\documentclass[12pt]{revtex4}
\usepackage{epsfig,amsbsy,bm,amssymb}
\usepackage{hyperref}

\newcommand{\isum}%
{\mathop{\hbox{$\displaystyle\sum\kern-13.2pt\int\kern1.5pt$}}}
\renewcommand{\r}{{\bm r}}

\newcommand{\p}{{\bm p}}

\newcommand{\bt}{\begin{tabular}}
\newcommand{\et}{\end{tabular}}

\newcommand{\eref}[1] {(\ref{#1})}
\newcommand{\Eref}[1] {Eq.~(\ref{#1})}

\newcommand{\Fref}[1] {Figure \ref{#1}}

\newcommand{\np}{\newpage}

\newcommand{\br}{\begin{eqnarray*}}
\newcommand{\er}{\end{eqnarray*}}
\newcommand{\ba}{\begin{eqnarray}}
\newcommand{\ea}{\end{eqnarray}}
\newcommand{\be}{\begin{equation}}
\newcommand{\ee}{\end{equation}}

\newcommand{\bp}{\begin{minipage}}
\newcommand{\ep}{\end{minipage}}

\begin{document}
\bibliographystyle{apsrev}

\title
{Evolution of the 
transverse photoelectron momentum distribution 
for atomic ionization 
driven by a laser pulse with varying
ellipticity}

\author{I. A. Ivanov }
\email{Igor.Ivanov@anu.edu.au}

\affiliation{Research School of Physical Sciences,
The Australian National University,
Canberra ACT 0200, Australia}

\date{\today}
\begin{abstract}
We consider the process of atomic
ionization driven by a laser pulse with varying ellipticity. 
We study distribution of the 
momenta of the
photoelectrons, ionized by a strong laser field, emitted in the
direction
perpendicular to the polarization plane (transverse distribution).
We show, that with changing laser pulse ellipticity, the 
transverse distribution evolves from the 
singular cusp-like
distribution for the close to linear polarization to the
smooth gaussian-like structure for the close to circular
polarization.
In the latter case, when the ellipticity parameter $\epsilon\to 1$
the strong-field approximation formula for the 
transverse momentum distribution  becomes quantitatively
correct.
\end{abstract}

\maketitle

\section{Introduction}

Tunneling theories of 
photo-ionization proved to be 
in the past (and still
are) of great importance in understanding atomic
or molecular ionization by strong infrared (IR) laser field. 
By tunneling theories we 
mean here a broad class of theories comprising original work
on the strong-field approximation (SFA) 
by Keldysh \cite{Keldysh64}, its modification known as
Keldysh-Faisal-Reiss (KFR) theory \cite{Faisal73,Reiss80}, 
subsequent developments 
\cite{tun1,tunr,BFHMF05,doublepeak5,coul3,coul5}, 
or quasistatic theories exploiting the fact, that IR field 
varies slowly in time 
\cite{adk,adk1}. 

The SFA in its original form \cite{Keldysh64} did not include
any interaction of ejected electron and parental ion. 
The importance of the effect of the Coulomb field
of the parental ion on the ionization process  has, however, 
long been realized.
Coulomb field of the residual ion 
was found responsible for such effects as low energy structures
in electron spectra \cite{coul6} or two dimensional momentum
distributions \cite{coul2}, 
Coulomb focusing \cite{coul9},
asymmetry in the spectra
of above threshold ionization spectra \cite{coul10}, formation of 
a cusp in the transverse  electron momentum distribution \cite{cusp2}.

It is the latter effect, the influence of the Coulomb field
of the parental ion on the transverse (i.e. perpendicular to the 
polarization plane of the driving pulse) electron momentum 
distribution, which will
interest us below. 
Study of the electron momentum  distributions
(both in longitudinal and
transverse directions) can shed light on fine details of the
strong-field ionization process \cite{cusp3,coul5}.

It has been found \cite{cusp2}, that for the case of the linearly polarized 
laser pulse, the transverse  electron momentum distribution 
exhibits a sharp cusp-like peak at zero transverse momentum.
For the 
case of the circularly polarized light, on the contrary,
this distribution was found not to deviate considerably from the 
gaussian-like structure predicted by the SFA \cite{coul7}.
We see, thus, two different forms (cusp-like and gaussian-like) of
the transverse  electron momentum distribution
for two 
limiting polarization states of the driving pulse. 
The aim
of the present work is to study transition between these two forms
in detail. 
We shall consider below atom
in the field of an elliptically polarized laser pulse, and 
consider evolution of the 
transverse  electron momentum distribution
with  varying ellipticity. We use hydrogen atom  as a model, 
atomic units are used 
throughout
the paper.

\section{Theory}

We solve the time-dependent Schr\"odinger equation (TDSE)
for a hydrogen atom:
\begin{equation}
i {\partial \Psi(\r) \over \partial t}=
\left(\hat H_{\rm atom} + \hat H_{\rm int}(t)\right)
\Psi(\r) \ .
\label{tdse}
\end{equation}

Operator  $\hat H_{\rm int}(t)$ in \Eref{tdse} 
describes interaction of the atom with the
EM field. We use 
velocity form for this operator:

\be
\hat H_{\rm int}(t) = {\bm A}(t)\cdot \hat{\bm p}\ ,
\label{gauge}
\ee

with 

\be
{\bm A(t)}=-\int_{0}^{t}{\bm E(\tau)}\ d\tau\\.
\label{vp}
\ee

The laser pulse is elliptically polarized (with ellipticity parameter 
$\epsilon$),
and  propagates along the
$z$-direction (assumed to be the quantization axis). Its field
components are:

\be
E_x= {E\over \sqrt{1+\epsilon^2}} f(t) \cos{\omega t} \ \ , \ \
E_y= {E\epsilon\over \sqrt{1+\epsilon^2}} f(t) \sin{\omega t} \ .
\label{ef}
\ee

In the calculations presented below we use $E=0.0534$ a.u.,
which corresponds to the 
intensity of $10^{14}$ W/cm$^2$ for the laser pulse described by 
\Eref{ef}. For the base frequency of the pulse we use $\omega=0.057$ a.u.
(corresponding to the wavelength of 790 nm). 
In \Eref{ef} the function  $f(t)= \sin^2(\pi t/ T_1)$ 
(here $T_1$ is a total pulse duration)
is a pulse envelope. 
We report below results 
for the pulse durations $T_1=4T$, and $T_1=10T$, 
where $T=2\pi/\omega$ is an optical cycle of the field \eref{ef}.

Initial state of the system is the hydrogen ground state.
To solve the TDSE we employ the strategy used in the works 
\cite{dstrong,circ6}.

Solution of the TDSE is represented as a partial waves series:
\be
\Psi({\bm r},t)=
\sum\limits_{l=0}^{L_{\rm max}} \sum\limits_{\mu=-l}^{l}
f_{l\mu}(r,t) Y_{l\mu}(\theta,\phi).
\label{basis}
\ee

The radial part of the TDSE is discretized on the grid with the stepsize
$\delta r=0.1$ a.u. in a box of the size $R_{\rm max}=400$ a.u. 
for the total pulse duration of 4 optical cycles, and 
$R_{\rm max}=1000$ a.u. for the pulse duration of 10 optical cycles.
In the calculations we used $L_{\rm max}=50$ in \Eref{basis}. A series
of routine checks ensuring that calculation is well converged with respect to 
variations of the parameters $\delta r$, $R_{\rm max}$, and
$L_{\rm max}$ has been performed.

Substitution of the expansion \eref{basis} into the TDSE gives us a 
system of coupled equations for the radial functions 
$f_{l\mu}(r,t)$, describing evolution of the system in time.
To solve this system 
we use the matrix
iteration method developed in \cite{velocity1}.
Ionization amplitudes $a(\p)$ 
are obtained by projecting solution of the 
TDSE at the end of the laser pulse
on the set of the ingoing scattering states 
$\psi^{(-)}_{\p}(\r)$ of hydrogen atom. 

In the present paper we are interested in the transverse electron
momentum distribution, 
describing probability to detect a photoelectron with a given value
of the momentum component $p_{\perp}$ perpendicular to the 
polarization plane. For the geometry we use
$p_{\perp}=p_z$. Transverse momentum distribution can be  obtained,
therefore, as:

\be
W(p_{\perp})=\int |a(\p)|^2\ dp_x\ dp_y
\label{wp}
\ee

The well-known SFA result for the transverse momentum distribution
in the case of the elliptically polarized laser pulse is 
a smooth gaussian form \cite{tunr}:

\be
W(p_{\perp})\propto
\exp
{
\left\{
    -{(2I)^{1/2}\sqrt{1+\epsilon^2}\over E} p_{\perp}^2 
\right\}
}\ ,
\label{wpg}
\ee

where $I$ is the ionization potential, and we employ the notation
used in \Eref{ef} for the laser field parameters.

This smooth distribution is in striking contrast to the cusp-like
structure observed in the experiment \cite{cusp2} for the 
ionization of noble gases by linearly polarized laser field 
(which corresponds to the case of $\epsilon=0$ in \Eref{wpg}).
This cusp-like structure has been attributed to the Coulomb effects,
which are, of course, neglected in the SFA.

In the next section we shall present results of a systematic study of the 
perpendicular momentum distributions for various values of the 
ellipticity parameter $\epsilon$. We shall see, that when ellipticity
parameter varies between $\epsilon=0$ (linear polarization) and 
$\epsilon=1$ (circular polarization), spectra evolve from the 
cusp-like spectra observed in \cite{cusp2} to the gaussian form 
predicted by the SFA and observed in \cite{coul7}.

\section{Results}

\begin{figure}[h]
\begin{tabular}{c}
\resizebox{120mm}{!}{\epsffile{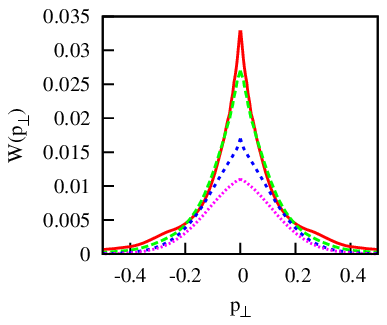}} \\
\resizebox{120mm}{!}{\epsffile{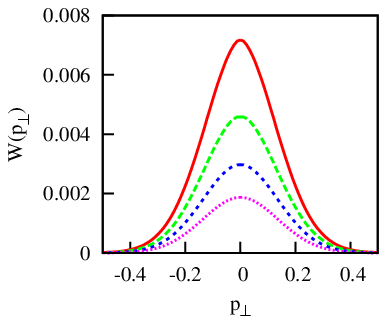}} \\
\end{tabular}
\caption{(Color online) Top panel: distribution $W(p_{\perp})$ for 
$\epsilon=0$ (solid (red) line), $\epsilon=0.2$  (dash (green)),
 $\epsilon=0.4$ (short dash (blue)),  $\epsilon=0.5$ (dots (magenta)).
Bottom panel: distribution $W(p_{\perp})$ for
$\epsilon=0.6$ (solid (red) line), $\epsilon=0.7$  (dash (green)),
 $\epsilon=0.8$ (short dash (blue)),  $\epsilon=1$ (dots (magenta)).
Pulse parameters: $E=0.0534$ a.u.,
$\omega=0.057$ a.u., pulse duration- four optical cycles.
\label{fig1}}
\end{figure}

\Fref{fig1} shows evolution of the distribution
$W(p_{\perp})$ with varying ellipticity parameter
$\epsilon$, when ionization is driven by a short laser pulse
with the duration $T_1=4T$ (four optical cycles).
A feature of the spectra clearly seen from the
\Fref{fig1} is the rapid decline of the transverse distribution
with the ellipticity parameter $\epsilon$. This decline
agrees with the \Eref{wpg}, which predicts exponential
dependence of transverse distribution on the ellipticity
parameter for a given field strength, which has the
same value $E=0.0534$ a.u. for all values of $\epsilon$.
Another feature clearly seen in the \Fref{fig1} is the gradual 
transition from the cusp-like structure in the spectra 
to the smooth gaussian form. As one can see, the 
cusp-like structure survives till $\epsilon\approx 0.5$.

\begin{figure}[h]
\begin{tabular}{cc}
\resizebox{70mm}{!}{\epsffile{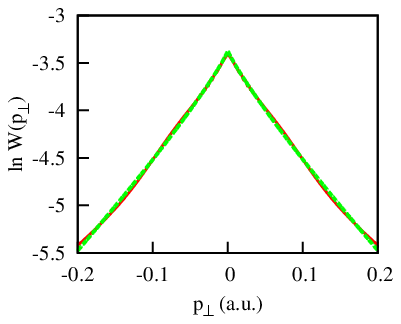}} &
\resizebox{70mm}{!}{\epsffile{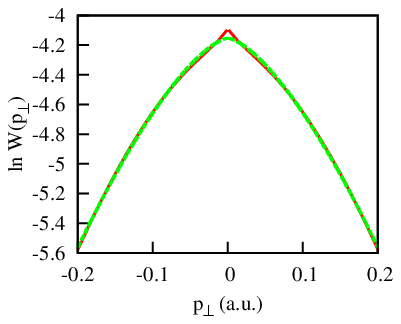}} \\
\resizebox{70mm}{!}{\epsffile{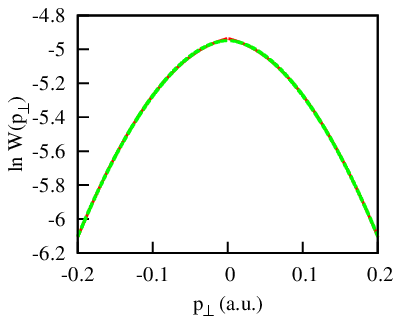}} &
\resizebox{70mm}{!}{\epsffile{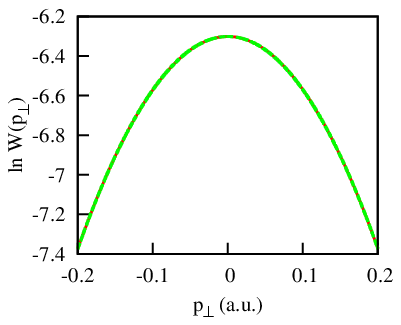}} \\
\end{tabular}
\caption{(Color online) Function $V(p_{\perp})=\ln{W(p_{\perp})}$
(red) solid line and results of the fit based on 
\Eref{exp} for
$\epsilon=0,0.4,0.6,1$ (left to right,
top to bottom).
Pulse parameters: $E=0.0534$ a.u.,
$\omega=0.057$ a.u., pulse duration- four optical cycles.
\label{fig2}}
\end{figure}

\begin{figure}[h]
\begin{tabular}{cc}
\resizebox{70mm}{!}{\epsffile{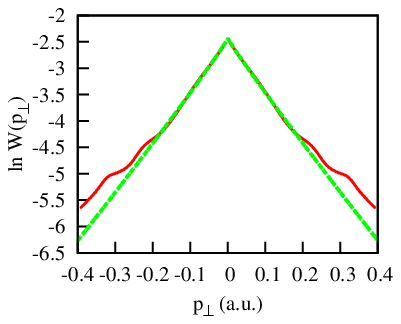}} &
\resizebox{70mm}{!}{\epsffile{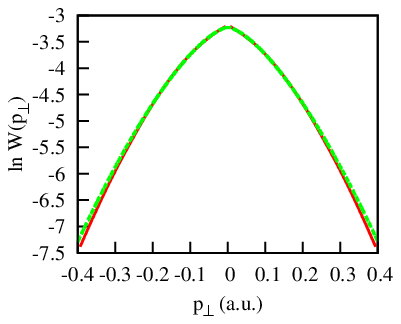}} \\
\resizebox{70mm}{!}{\epsffile{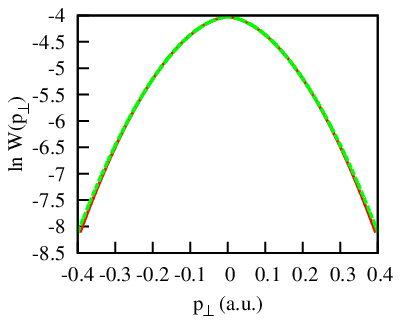}} &
\resizebox{70mm}{!}{\epsffile{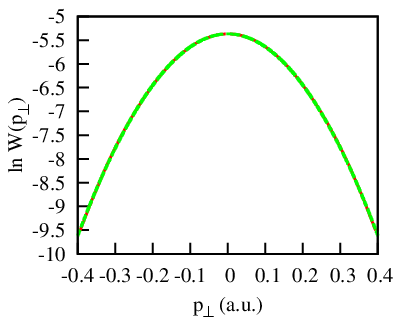}} \\
\end{tabular}
\caption{(Color online) Function $V(p_{\perp})=\ln{W(p_{\perp})}$
(red) solid line and results of the fit based on
\Eref{exp} 
for 
$\epsilon=0,0.4,0.6,1$ (left to right,
top to bottom).
Pulse parameters: $E=0.0534$ a.u.,
$\omega=0.057$ a.u., pulse duration- ten optical cycles.
\label{fig3}}
\end{figure}

To gain better insight into this interesting phenomenon 
we show in \Fref{fig2}, \Fref{fig3} the function 
$V(p_{\perp})=\log{W(p_{\perp})}$ 
for a narrow interval in the vicinity of the
singular point $p_{\perp}=0$. For the distribution $W(p_{\perp})$
to have a cusp at $p_{\perp}=0$, $V(p_{\perp})$ must clearly be singular
there. The nature of this singularity can be guessed from the 
character of the singularity of  $W(p_{\perp})$
seen in \Fref{fig1} for small values of the ellipticity parameter
$\epsilon$. One can see, that 
$W(p_{\perp})$ is continuous at $p_{\perp}=0$, but may have 
infinite derivatives
there. Clearly, $V(p_{\perp})$ must possess the same properties. It is 
natural, therefore, to suggest, that in the vicinity of the 
singular point $p_{\perp}=0$ function 
$V(p_{\perp})$
can be represented as:

\be
V(p_{\perp})= B+A|p_{\perp}|^\alpha\ .
\label{exp}
\ee

We may look at this equation as a tentative
expression for the first two terms of an 
expansion of $V(p_{\perp})$ in the vicinity of the singular point. 
We perform next a series of the least square fits,
fitting the function  $V(p_{\perp})$ using \Eref{exp}, and 
considering the coefficients  $A$, $B$, $\alpha$ in this equation
as fitting parameters.

Results are shown in \Fref{fig2} and \Fref{fig3}. One can see,
that \Eref{exp} approximates
the function $V(p_{\perp})$ in the vicinity of the singular point
fairly accurately. 
The quality of the fit 
does not depend appreciably on the interval of data we use
as an input for the fitting procedure, as long as this interval is 
sufficiently narrow (results shown in \Fref{fig2} and \Fref{fig3}
have been obtained using the interval $p_{\perp}\in (-0.2,0.2)$
for the fitting procedure). The accuracy of the fit shown in
\Fref{fig2},\Fref{fig3} makes the conjecture that 
\Eref{exp} indeed describes faithfully the behavior of 
$V(p_{\perp})$ in the vicinity of the singular point plausible
enough. Further support in favor of the validity of this 
conjecture is found at a closer inspection of the 
coefficients of the fitting formula \eref{exp}, which the fitting
procedure gave us.

Fitting parameters $A$ and $\alpha$ describing the character of the 
singular point are of most interest to us. Their dependence on the 
ellipticity parameter $\epsilon$ is shown in \Fref{fig4}.
For the nearly circular polarization parameter $\alpha\approx 2$,
as prescribed by the tunneling
\Eref{wpg}. Moreover, 
parameter $A$, as predicted by the tunneling equation, 
should approach the 
value $A=-26.48$ for $\epsilon=1$. We see thus, that for the nearly
circular polarization tunneling \Eref{wpg} describes the electron spectrum 
fairly accurately. With decreasing values of the ellipticity parameter
deviation from the gaussian-like behavior due to the Coulomb effects 
becomes more and more significant. 

This behavior can be understood if we take into account, that the
angular momentum composition of the wave function changes considerably with
the ellipticity parameter. If we rewrite \Eref{basis} as a sum of 
$\Psi_l({\bm r},t)$, where each $\Psi_l({\bm r},t)$ 
includes only spherical harmonics of
rank $l$, than distribution of the squared norms  
$N_l=|\Psi_l|^2$ can be used to
characterize the angular momentum composition of the wavefunction. 
Since absorption of a photon from the circularly
polarized wave increases magnetic quantum number by one unit, we can expect,
that this distribution will be shifted towards larger $l$ for pulse
polarization approaching a circular one. That reasoning is illustrated in 
\Fref{fig5}, where distributions of the norms $N_l$ for the
wavefunction after the end of the laser pulse are shown
for the cases of $\epsilon=0$, and $\epsilon=0.8$. We subtracted 
contribution of the ground state to the $l=0$ component.
The expected shift towards larger $l-$ values is clearly seen in the 
\Fref{fig5}. This shift towards larger angular momenta means much 
higher centrifugal barrier, and consequently larger electron-parental
ion
separation for the 
close to circular polarization, which should diminish the role of 
the Coulomb effects.

The role of the high angular momenta can be further clarified, if
we consider 
transverse distributions obtained by putting in \Eref{basis} all
components of 
the wavefunction with angular momenta less than certain value
$L_{\rm min}$,
to zero. This means, that projecting the solution of the TDSE
on the ingoing scattering states of the hydrogen atom, we 
ignore all Coulomb continuum wavefunctions with 
$l<L_{\rm min}$.

\Fref{fig6} shows 
the transverse distributions computed for different values of the 
parameter $L_{\rm min}$ for the case of the linearly polarized laser pulse.
One can see, that if we leave in the wavefunction only the components
with angular momenta $l\ge 10$ the cusp structure disappears and 
we obtain gaussian transverse distribution. It was suggested 
in \cite{cusp2}, that for the case of the linear polarization,
cusp originates
from the singularity of the Coulomb continuum 
wavefunction at zero energy. Absence of the cusp in \Fref{fig6}
for $L_{\rm min}=10$ suggests, that if we use for the projection
operation only the Coulomb continuum wavefunctions with 
high angular momenta, this singularity somehow disappears, or at least
becomes less visible.

A plausible explanation for this fact can be obtained, if we 
consider in more  detail the projection operation
we perform to find the ionization probabilities. 
This projection operation, as we mentioned above, 
relies on the calculation of the overlap integrals between the TDSE
solution and the Coulomb continuum wavefunctions. The Coulomb continuum
wavefunction is an object
notoriously singular at the point 
$E=0$. 
Overlap integrals  
naturally inherit this singularity,
which, as was suggested in \cite{cusp2}, produces cusp in the 
momentum distribution.

The character of the singularity
inherited from the Coulomb continuum wavefunctions may,
however, be different for 
different angular momenta $l$. 
One can put forwards arguments showing, that this singularity 
becomes milder
with increasing $l$. The well-known asymptotic expression
for the radial  Coulomb continuum wavefunction for $r\to 0$,
$E\to 0$ is
(we use energy normalization here) \cite{LL3}:

\be
R_{El}(r) \approx {2^{l+1}\over(2l+1)!} r^{l+1} \ 
\qquad r\to 0, E\to 0
\label{ask1}
\ee

This expression,
considered as a function of energy, is of course,
non-singular. 
The TDSE solution
we obtain, considered as a function of spatial variables,
has some finite spatial extension $a$. 
It is a known property of the Coulomb wavefunctions \cite{abr},
that for any $a$, the small-$r$ asymptotic expression \eref{ask1} 
faithfully reproduces $R_{El}(a)$ if angular momentum 
$l$ is high enough. For such $l$, substitution of the asymptotic
form \eref{ask1} into the overlap integral is legitimate, and
we obtain a result which is non-singular in energy. 
We can see, therefore, 
that
the overlap integrals between the TDSE solution and the 
continuum Coulomb wavefunctions become less and less singular
with increasing  $l$. This, we believe, is the mechanism,
which effectively removes the singularity in the overlap integrals
and hence in the momentum distributions, if only high angular 
momenta are important in the expansion 
\eref{basis} of the
wavefunction. This, in particular, is the case of the atomic
ionization by a laser pulse 
with close to circular
polarization.

\section{Conclusion}

We studied evolution of the transverse momentum
distribution $W(p_{\perp})$,
describing distribution of electron momenta in the direction 
perpendicular to the polarization plane, with the change of the
ellipticity parameter. We saw the gradual change of the 
character of this distribution from the singular cusp-like
distribution for the close to linear polarization, to the 
smooth gaussian-like structure for the close to circular
polarization state. 

\begin{figure}[h]
\begin{tabular}{cc}
\resizebox{70mm}{!}{\epsffile{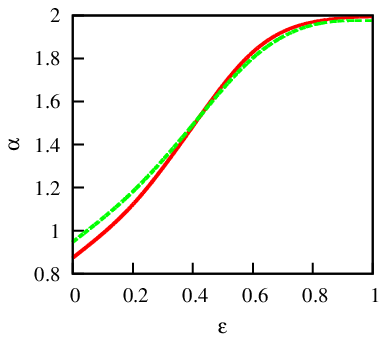}} &
\resizebox{70mm}{!}{\epsffile{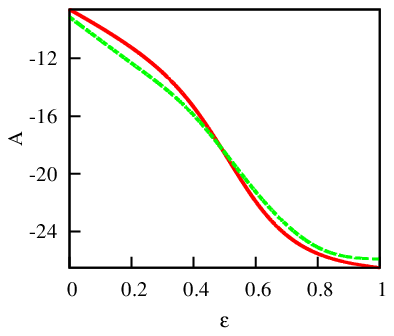}} \\
\end{tabular}
\caption{(Color online) Fitting parameters
$A$ and $\alpha$ in \Eref{exp} as functions of the
ellipticity parameter $\epsilon$. 
Pulse duration: 4 optical cycles:
(red) solid line; 10 optical cycles (green) dash.
Other pulse parameters: $E=0.0534$ a.u.,
$\omega=0.057$ a.u.
\label{fig4}}
\end{figure}

\begin{figure}[h]
\begin{tabular}{cc}
\resizebox{70mm}{!}{\epsffile{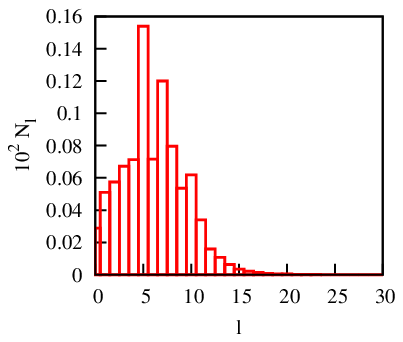}} &
\resizebox{70mm}{!}{\epsffile{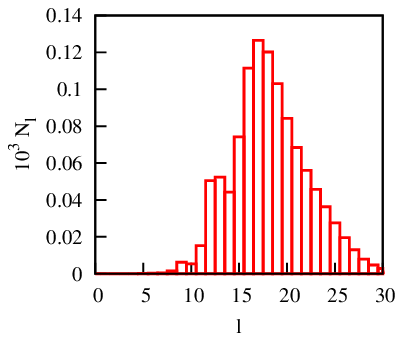}} \\
\end{tabular}
\caption{(Color online) Angular momentum 
distributions for $\epsilon=0$ (left panel) and
$\epsilon=0.8$ (right panel).
Pulse parameters: $E=0.0534$ a.u.,
$\omega=0.057$ a.u., pulse duration- four optical cycles.
\label{fig5}}
\end{figure}

\begin{figure}[h]
\begin{tabular}{c}
\resizebox{120mm}{!}{\epsffile{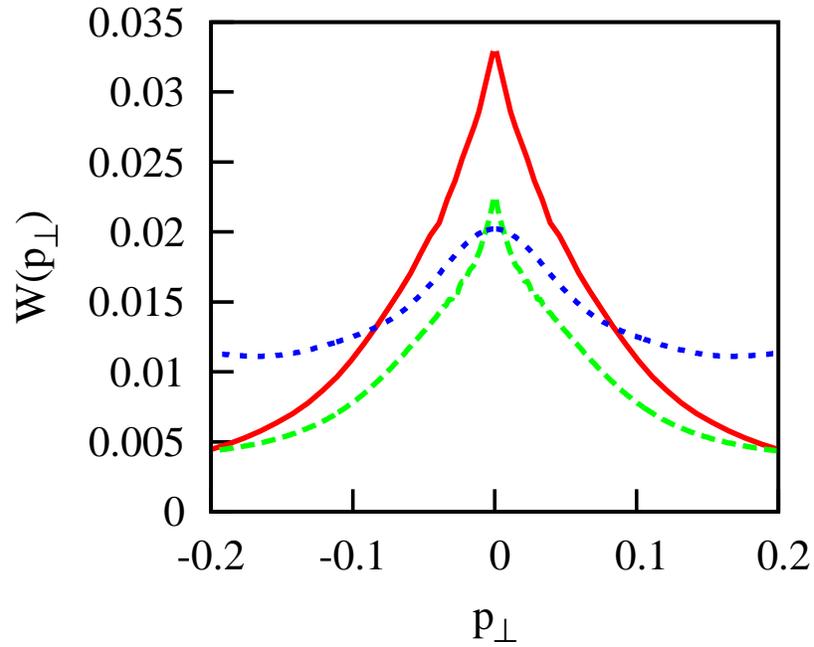}} 
\end{tabular}
\caption{(Color online)  
Distribution $W(p_{\perp})$ .
Projection on the Coulomb scattering states with
$L_{\rm min}=0$ (solid (red) line), $L_{\rm min}=5$  (dash (green)),
$L_{\rm min}=10$ (short dash (blue)). Distribution for $L_{\rm min}=10$
is scaled by a factor of 10 for better visibility.
Pulse parameters: $\epsilon=0$, $E=0.0534$ a.u.,
$\omega=0.057$ a.u., pulse duration- four optical cycles.
\label{fig6}}
\end{figure}

In the latter case, when the ellipticity parameter $\epsilon\to 1$
we see, that the tunneling formula \eref{wpg} becomes quantitatively
correct. It does not necessarily mean, that Coulomb effects
are completely absent for the circular polarization.
It is hardly possible,
indeed, that a singular point of a
generic function depending on a parameter
(ellipticity parameter in the present case) can disappear completely
when the parameter gradually changes. We would rather be inclined to
believe, that the singularity simply becomes much more difficult to see.
Indeed, the \Fref{fig4}
shows, that exponent $\alpha$ in \Eref{exp} is less than 1 for 
$\epsilon=0$. This means, that $W(p_{\perp})$ has infinite first
derivative at $p_{\perp}=0$. At the value of the 
ellipticity parameter $\epsilon\approx 0.1$,
first derivative of $W(p_{\perp})$
ceases to be infinite at $p_{\perp}=0$. It is the second derivative
of $W(p_{\perp})$ which becomes infinite at this point. 
The singularity, therefore, becomes milder but does not go away
completely. We believe, this is what might happen for 
$\epsilon=1$,
some higher order derivatives of  $W(p_{\perp})$
at the point $p_{\perp}=0$
are still infinite, which is, of course, very difficult to
establish in numerical calculation.

From the practical point of view, the Coulomb effects in 
the distribution $W(p_{\perp})$ of the
electron momenta in the direction
perpendicular to the polarization plane become hardly important 
for the ellipticity parameter as large as $0.8$. As we have seen,
the tunneling formula \eref{wpg} works quite reliably for 
such degree of polarization. The reason for this diminishing of
the role of the Coulomb effects can be attributed, we believe, to
the mechanism which we described above, and which effectively
removes the singularity from the overlaps integrals for the 
close to circular polarization of the laser pulse.

\subsection*{ACKNOWLEDGMENT}

Author acknowledges support of the Australian Research Council in
the form of the Discovery grant DP120101805.  Author is grateful to
Prof. A.S.Kheifets for the stimulating discussions and to a
referee for a valuable suggestion.
Resources of the National
Computational Infrastructure (NCI) Facility were employed.

\np


\end{document}